\documentclass[twocolumn,showkeywords,preprintnumbers,amsmath,amssymb]{revtex4}

\usepackage{graphicx}
\usepackage{dcolumn}
\usepackage{bm}

\begin{document}

\title{Semantics of Information}

\author{Daegene Song}

\affiliation{%
Department of Management Information Systems, Chungbuk National University, Cheongju, Chungbuk 28644, Korea
}%

\date{\today}

\begin{abstract}
Due to the self-referencing aspect, consciousness is placed in a unique non-computable position among natural phenomena. Non-computable consciousness was previously analyzed on the basis of self-referential cyclical time. This paper extends the cyclical model of vacuum observation and posits that choice, or the experience of reality, may be expressed as the initial part of the self-referencing loop, while the conscious awareness of the experience is the other part of the loop. In particular, the inseparability of the two sides of the loop is established through the cyclical time process, which bears a resemblance to Heidegger's analysis of existence. The cyclical looping model is also discussed in terms of Wittgenstein's analysis of language as attaching semantic meaning, or continuous or infinite conscious awareness, to physical reality. We also discuss the proposed model of subjectivity and cyclical time - as opposed to objectivity and linear time - which may be considered similar to Hebrew thought.

\end{abstract}

\maketitle

\section{Introduction}
With recent advancements in the field of artificial intelligence (AI), much of what was formally unthinkable has been achieved by a computing machine, such as winning the quiz show Jeopardy \cite{jeopardy} or beating human chess or Go champions \cite{chess,go}. The inquiry then arises as to how far the development of AI can proceed. Indeed, one of the intriguing questions in this regard is whether it would ever be possible for a computer to have a conscious ego, a term also known as strong AI. There are divided opinions \cite{kurzweil,chinese} regarding this question. For example, Penrose argued \cite{penrose} that the G\"odel-type non-algorithmic process, i.e., the non-computable element, may be present in consciousness. In \cite{song2007}, it was found, as suggested by Penrose, that there does exist a non-computable element in consciousness, which follows the logic of self-referencing.

A number of studies have suggested \cite{deutsch, thooft,wolfram, schmidhuber,lloyd2006, vedral,zizzi} a close relation between nature and computation. If we accept this analogy, then it is rather odd that consciousness, which is considered part of nature, has a non-computable element. One attitude toward this awkward situation is to accept that consciousness is simply different from the rest of nature. The other approach may be to examine whether there is something amiss with the way in which the rest of nature is considered and whether the rest of nature could be modified such that it is consistent with consciousness. The intention of this paper is to accomplish the latter.

\section{Temporal}
In general, an observation in quantum theory involves a state vector and a reference frame. In the case of self-observation, the state vector, which is the object, is the same as the reference frame or the observable. In that case, the active and passive approaches are not generally equivalent \cite{song2007}, therefore, self-observation is not computable in a linear time model. 

A cyclical time model was recently introduced to analyze non-computable consciousness from a different perspective. Although circulating time (see \cite{reynolds} for a review), as opposed to the familiar linear time, seems counter-intuitive, it is useful for understanding the strange phenomenon of consciousness. The time looping process can be considered in the context of self-referencing such as the following liar's paradox:
\begin{enumerate}
\item The following sentence is true
\item The previous sentence was false
\end{enumerate}
It can be seen that this two-sentence version of self-referencing is equivalent to the single-sentence liar's paradox. The first sentence, or the initial part of the loop, refers to a future event, while the second part of the loop refers to the past; therefore, the logic is based on circulating time. 

\begin{figure}
\begin{center}
{\includegraphics[scale=.75]{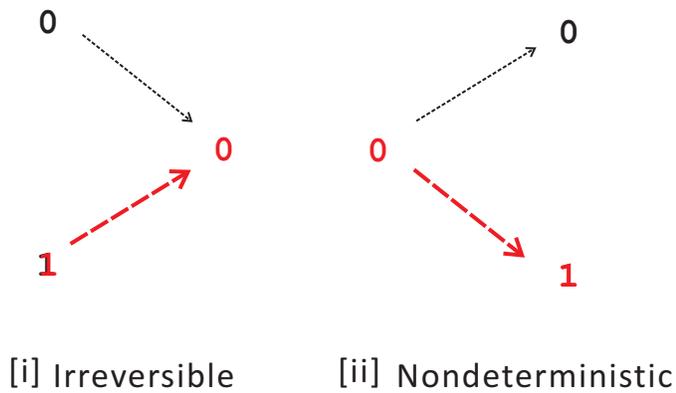}}

\end{center}
\caption{ [i] The irreversible computation \cite{landauer} refers to the process whereby either input 0 or 1 is set to 0, i.e., erasing one bit of information. The nondeterministic computation [ii] is the time-reversal process of [i].     }
\label{irre}\end{figure}

A similar analysis may be applied to the physical version of self-referencing, or consciousness, as follows:
\begin{enumerate}
\item The observer makes an observation
\item  The observed was the observer
\end{enumerate}
While the first part of the looping is time-forwarding, the closing part has to do with time moving backward. In \cite{song2017d}, the cyclical looping of consciousness was applied to the cosmological constant problem, which concerns the disparity between the theoretical and observation values of vacuum energy \cite{carroll}. The vacuum observation may be seen as self-observation in the sense that the conscious observer is observing her or his own reference frame of energy states. Based on the two-sentence, self-referencing argument, the vacuum observation in the cyclical model was discussed as follows:
\begin{enumerate}
\item The classical reference frame performs irreversible computation (Fig. \ref{irre} [i])
\item The quantum reference frame performs nondeterministic computation (Fig. \ref{irre} [ii])
\end{enumerate}
Thus, while the classical experience is done in a time-forward manner, the second part of the loop involves the quantum reference frame as evolving backward in time. In \cite{song2017d}, it was shown that the discrepancy in the cosmological constant problem should correspond to the difference between the energy resulting from the classical computation moving in a time-forward manner and the energy resulting from the quantum evolution - or the Dirac-type negative sea of consciousness \cite{song2017a}.

\section{Inseparable}
\begin{figure}
\begin{center}
{\includegraphics[scale=.65]{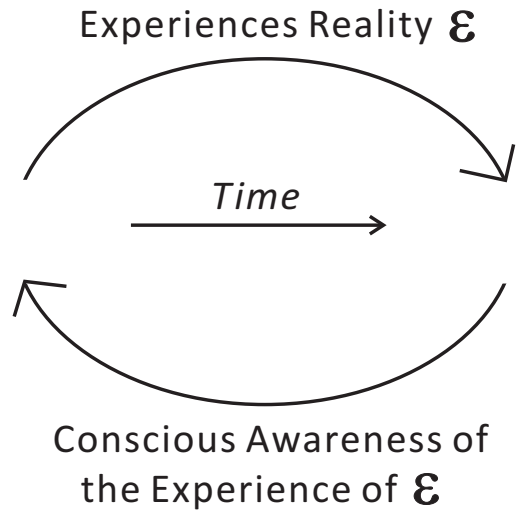}}

\end{center}
\caption{ The cyclical looping process of reality and conscious experience: The first part of the loop corresponds to the experience of reality, i.e., the classical choice being made, while the closing part is the conscious realization of the experience of reality through a nondeterministic quantum computation.   
    }
\label{loop}\end{figure}

We now wish to extend the case of the cyclical looping model of vacuum observation to the process of experiencing physical reality. Reality refers to something that can be directly experienced in classical space, such as energy, mass, etc. In \cite{landauer}, Landauer showed that the process of irreversible computation erasing a bit yields energy, $E_0 = kT\ln 2$. If we assume that 
\begin{equation}
{\mathcal{E}}=N\cdot E_0
\label{reality}\end{equation}
then the process of experiencing the reality of ${\mathcal{E}}$ may be considered with the following looping process: 
\begin{enumerate}
\item Experiences the reality of ${\mathcal{E}}$ through classical irreversible computation
\item Conscious awareness of the experience of ${\mathcal{E}}$ through quantum nondeterministic computation
\end{enumerate}
The first part of the loop corresponds to the classical reference frame of performing an irreversible computation, i.e., the classical choice is being made (Fig. \ref{loop}). Since the irreversible computation is associated with energy, this implies that the choice made by the observer corresponds to the experience of energy ${\mathcal{E}}$. Note that there have been numerous discussions suggesting the relation between symmetry and physical reality \cite{jacobson,padmanabhan,verlinde}. The second part of the loop corresponds to the conscious understanding or awareness of the experience of reality ${\mathcal{E}}$ through the evolution of the quantum reference frame.

The proposed model is similar to the philosophical analysis of existentialism. For instance, Heidegger noted  \cite{heidegger} that the reality of the observed and the conscious observer are not separable; as such, he advanced the notion of {\it{dasein}}, i.e., being-in-the-world. In particular, he addressed the being's nature with time, i.e., past, present, and future together compose the being. The above model (Fig. \ref{loop}) shows the experience of physical reality, i.e., the experience of choice, as being inseparable from the conscious realization of the physical experience through the cyclical process of time.

\section{Language}
The argument in \cite{hardy} was that classical probability theory may be considered equivalent to quantum theory, except for the aspect of continuity. Moreover, \cite{song2016} argued that one can exploit this idea such that classical probability theory can be considered to involve a classical choice made by the observer, whereas continuity is associated with the quantum part. Thus, while the experience of physical reality is discrete or finite, conscious awareness, which corresponds to quantum evolution, is continuous. Therefore, one may consider that the semantics of the finite, i.e., the experience of reality, are associated with continuity or infinity, i.e., conscious realization.

\begin{figure}
\begin{center}
{\includegraphics[scale=.6]{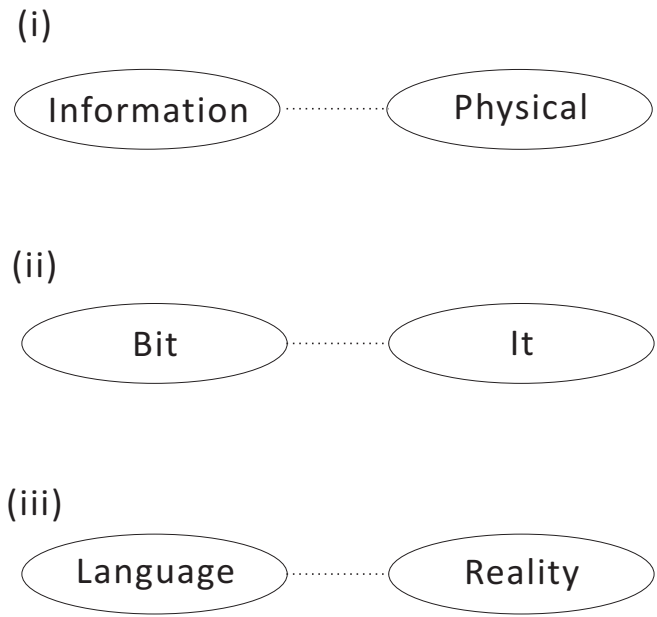}}

\end{center}
\caption{Various studies have examined the close ties between reality and information: (i) Landauer argued that information is physical, while (ii) Wheeler coined the term {\it{it from bit}}; (iii) in philosophy, Wittgenstein attempted to explain the limits of language based on the relationship between language and reality.   
    }
\label{bit}\end{figure}

\begin{figure}
\begin{center}
{\includegraphics[scale=.7]{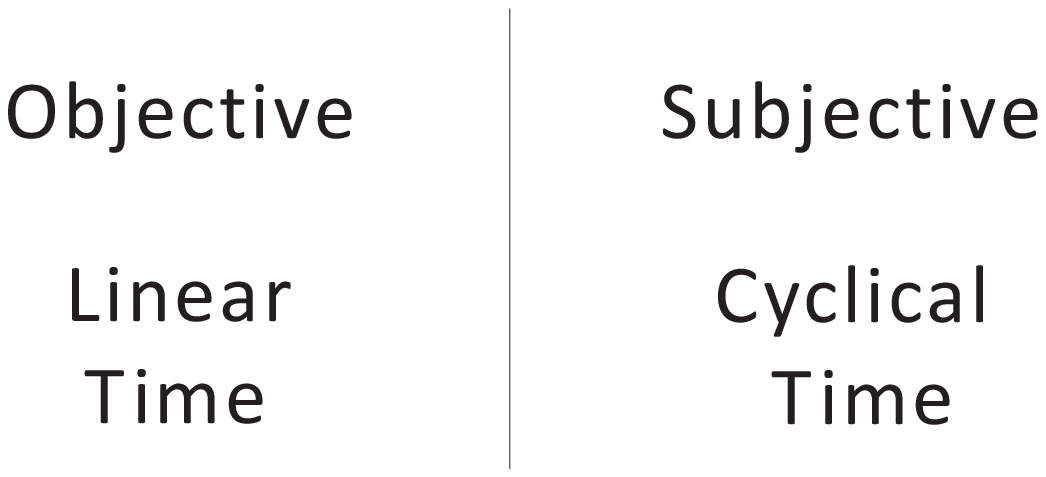}}

\end{center}
\caption{ While linear time and objective reality are widely used in traditional studies, subjectivity and cyclical time are often found in Hebrew thought \cite{boman}. The cyclical model proposed in this paper bears a resemblance to the latter case.   
     }
\label{thought}\end{figure}

As noted earlier, there have been various discussions regarding the relation between information and reality (Fig. \ref{bit}).  In \cite{landauer1}, Landauer explained the physical nature of information, for example, with the following quote: 
\begin{center}
{\it{Computation is inevitably done  with real physical degrees of freedom, obeying the laws of physics, and using parts available in our actual physical universe.}}
\end{center}
In \cite{wheeler}, Wheeler also speculated the close ties between reality and information, with the now well-known phrase {\it{it from bit}}. From a philosophical standpoint, Wittgenstein examined the relationship between language and reality \cite{wittgenstein,wittgenstein1}, noting that language plays the role of attaching semantics to reality. Indeed, this attachment is similar to the role of cyclical time in the sense that the looping process interconnects conscious awareness with reality, as shown in Fig. \ref{loop}.

It is remarkable that one is able to communicate or learn the meaning of infinity with only a finite array of binary bits. This suggests that the continuity or infinite aspect may not be something that is transferred or learned, i.e., the semantic aspect of language ought to be innate and shared. Indeed, as indicated earlier \cite{song2016}, this continuous or infinite side of consciousness may be associated with universal grammar, which argues that all languages have a universal structure that is not learned but is innate in human nature \cite{chomsky1,chomsky2,cook}. A similar argument of the shared Dirac-type negative sea of consciousness was discussed \cite{song2017c} in terms of entanglement and nonlocality.

\section{Remarks}
In \cite{song2012}, the suggestion of subjective reality was drawn from the argument of consciousness, that is, only in the case of consciousness are the object and the observer the same. This phenomenon is unique, considering that the observation or measurement results from the relative difference between the object being observed and the observer. Consequently, it has been suggested that because of consciousness, one should stop considering the observer and the object as separate entities.

Conventional thought has often distinguished between the observer and the observed, mind and physical, etc. This approach is consistent with the scientific endeavor, at least until the arrival of quantum theory, in seeking an objective reality. However, this has not been the case in Hebrew thought, i.e., the description of subjective reality has often been used  \cite{boman}. The proposed model \cite{song2012}, which is based on quantum theory and consciousness, bears a resemblance to the Hebrew conception of reality. In particular, the concept of cyclical time, through which the subjective reality has been placed on firmer ground, as discussed in this paper, is often found in the Hebrew language as well (Fig. \ref{thought}).

One of the mysteries in physics has been the apparent discrepancy between classical and quantum theories. This is because quantum theory involves many elements not seen in classical physics \cite{peres}, such as wave/particle duality, superposition, entanglement, nonlocality, etc. There is also a historical reason for this confusion; quantum theory was introduced in the twentieth century, and its descriptions of certain behaviors were presumed to replace many previously considered classical behaviors. This explains why one of the largest efforts in the theoretical physics research community is now being directed toward trying to come up with a quantum version of classical theory, namely, gravitation. However, contrary to the common perception, the classical aspect may be an integral part of the whole picture. Indeed, the presented cyclical model shows the manner in which the classical and quantum aspects are interconnected through time.


\end{document}